\documentclass[%
 reprint,
 amsmath,amssymb,
 aps,
]{revtex4-1}

\usepackage{graphicx}
\usepackage{dcolumn}
\usepackage{bm}
\usepackage{subcaption}

\def\RE{{\rm Re}} 
\def\IM{{\rm Im}}

\def\pv{{\bf{p}}}

\def\kv{{\bf{k}}}

\def\ntil{\tilde{\bf{n}}}

\def\barr{\begin{eqnarray}}
\def\earr{\end{eqnarray}}
\def\kt{     { \bf{k}_{\rm T}  }    }

\def\kpt{  \kv'_{\rm T} }
\def\ktkt{\kv^2_{\rm T}}
\def\kptkpt{{\kv'}^2_{\rm T}}

\def\pt{     { \bf{p}_{\rm T}  }    }
\def\ptpt{\pv^2_{\rm T}}
\def\bt {b\T}
\def\bl {b\L}
\def\T{_{\rm T}}
\def\L{_{\rm L}}
\def\A{_{\rm A}}
\def\B{_{\rm B}}

\def\QA{q_{\rm A}}

\def\QbB{\bar{q}_{\rm B}}

\def\A{_{\rm A}}
\def\B{_{\rm B}}

\def\be{\begin{equation}}
\def\ee{\end{equation}}

\begin{document}

\preprint{APS/123-QED}

\title{A simplified recursive ${}^3P_0$ model for the fragmentation of polarized quarks}

\author{A. Kerbizi$^{\, 1}$, X. Artru$^{\, 2}$, Z. Belghobsi$^{\, 3}$ and A. Martin$^{\, 1}$}
\affiliation{\\ $^{1}$ {\small INFN Sezione di Trieste and Dipartimento di Fisica, Universit\`a degli Studi di Trieste,}\\
{\small Via Valerio 2, 34127 Trieste, Italy}\\
$^{2}$ {\small Univ. Lyon, Universit\'e Lyon 1, CNRS,}
{\small Institut de Physique Nucl\'eaire de Lyon, 69622 Villeurbanne, France}
\\$^{3}$ {\small Laboratoire de Physique Th\'eorique, Facult\'e des Sciences Exactes et de l'Informatique,}\\ 
{\small Universit\'e Mohammed Seddik Ben Yahia,}\\
{\small B.P. 98 Ouled Aissa, 18000 Jijel, Algeria}\\
}
 
\date{\today}

\begin{abstract}
We revisit our recursive model for the fragmentation of polarized quarks based on the string+${}^3P_0$ mechanism of $q\bar{q}$ pair creation. We make a different choice for one input function of the model that simplifies the implementation in a Monte Carlo program. No new parameters are introduced, and the relevant results are the same apart from the suppression of the spin-independent correlations between successive quarks. In addition, the present version is more suitable for an interface with external event generators.
The theoretical aspects and the simulation results obtained with a stand alone program are discussed in detail and compared with those of the previous version of the model.

\end{abstract}

\maketitle


\section{Introduction}

The theoretical description of high energy collisions like $e^+e^-$ annihilations, lepton-nucleon Deep Inelastic Scattering (DIS) and inelastic $pp$ scattering involves factorization theorems which separate the sub-processes calculable in perturbative QCD from the non-perturbative ones. For the semi-inclusive processes where at least one hadron is detected in the final state, the knowledge of fragmentation functions (FFs) is needed. They are universal functions which describe how the coloured quarks and gluons transform into observable hadrons and cannot be calculated perturbatively.
This issue has been tackled through models, for instance, inspired from field theory or of multi-production type \cite{Field-Feynman,Artru-Mennessier,Lund1983}.

\begin{figure}[htb]
  \centering
    \includegraphics[width=0.36\textwidth]{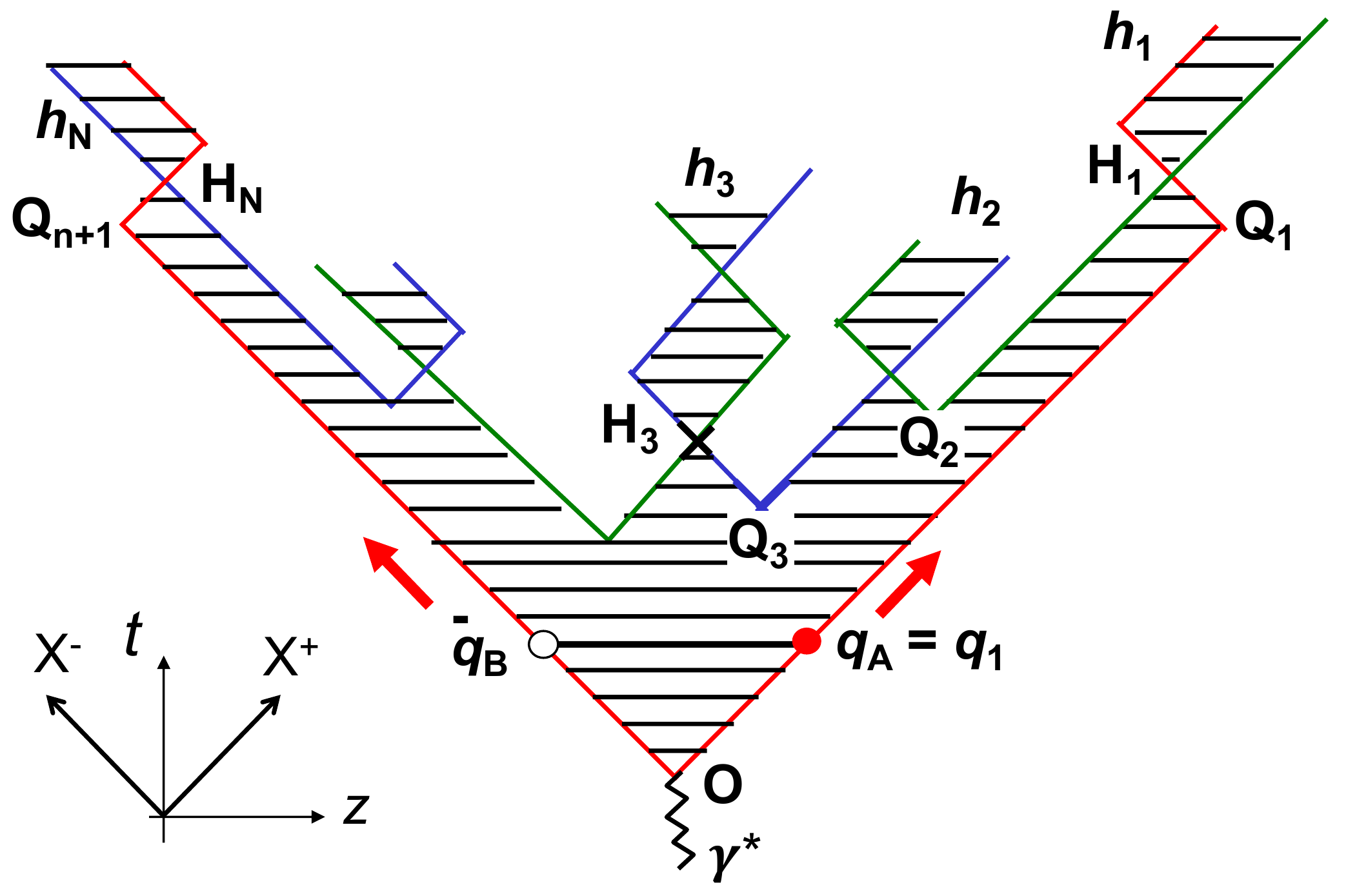}
  \caption{Space-time history of the hadronization process of a $q_{\rm{A}}\bar{q}_{\rm{B}}$ system without gluons produced in $e^+e^-$ annihilation. $Q_2,Q_3,\dots$ are the string breaking points whereas $H_1,H_2,\dots$ are the hadron emission points.}\label{fig:space_time_history}
\end{figure}

Within the latter class of models, the most successful one is the Symmetric Lund Model (SLM) \cite{Lund1983}, where the interaction among color charges is treated as a relativistic string which decays by a tunneling process into smaller string pieces through the creation of $q\bar{q}$ pairs in the string world-sheet. Such a chain is depicted in Fig. \ref{fig:space_time_history} for an initial quark-antiquark pair $q_{\rm{A}} \bar{q}_{\rm{B}}$ that hadronizes into mesons. Tunneling of diquarks can account for baryon production. The SLM is symmetric under the reversal of the quark line, namely the hadronization process can be viewed to occur from the $q\A$ side to the $\bar{q}\B$ side or from $\bar{q}\B$ to $q\A$ with the same probability. This symmetry will be referred to as the LR symmetry, ``LR'' standing for \textit{left-right} according to \cite{Lund1983} or less subjectively for ``Line Reversal''. This requirement is a strong and important constraint on the form of the splitting function of the SLM.

The SLM has been implemented in Monte Carlo event generators like PYTHIA \cite{pythia8} which is successfull in the description of experimental data from unpolarized reactions. However it does not incorporate polarization effects.

By now it is well estabilished that the quark polarization produces important effects in $e^+e^-$ annihilation \cite{belle-spin-asymmetries,besIII,babar} and in polarized reactions like semi-inclusive DIS (SIDIS) \cite{hermes-ssa,COMPASS-collins-sivers,jlab-SSA} where large transverse spin asymmetries have been observed for single hadrons and hadron pairs in the same jet \cite{interplay,compass-dihadron}. Particularly relevant is the Collins effect \cite{collins}, an asymmetry in the azimuthal spectrum of hadrons produced in the fragmentation of a transversely polarized quark. It is described by the Collins transverse momentum dependent fragmentation function (TMD FF), a non-perturbative and universal function which in SIDIS is coupled to the quark transversity distribution resulting in an observed azimuthal modulation of the hadron in $\sin(\phi_h+\phi_{\textbf{S}})$ in the $\gamma^*$-nucleon frame, where $\phi_{\textbf{S}}$ is the azimuthal angle of the nucleons transverse polarization about the $\gamma^*$-nucleon collision axis. This asymmetry is then used as an observable to access transversity \cite{Anselmino2013,Bacchetta2013,M.B.B}.

Attempts for the inclusion of the quark spin in the fragmentation process have been made in the past. In particular the model of Ref.\cite{DS09,DS11,DS13} is an extention of the SLM where the $q\bar{q}$ pairs at string breaking are produced in the ${}^3P_0$ state. An alternative model based on a field theoretical approach has been presented in Ref. \cite{Mate}.

In the $string+{}^3P_0$ model the quark spin is encoded in $2\times 2$ density matrices and treated with rules that preserve the LR symmetry. We have recently implemented the general $string+{}^3P_0$ model in a stand alone MC program \cite{kerbizi-2018} which simulates the fragmentation of a quark (or anti-quark) with arbitrary polarization into pseudoscalar mesons. The comparison of the resulting Collins and dihadron asymmetries with experimental data from SIDIS and $e^+e^-$ are very promising \cite{kerbizi-2018}.


The present work is based on the previous study of Ref. \cite{kerbizi-2018} and a simpler choice of one input function of the $string+{}^3P_0$ model is done. The model is completely LR symmetric and is characterized by a splitting function without dynamical spin-independent correlations between the transverse momenta of two successive quarks \cite{artru-belghobsi-essma}. It leads to simpler simulation codes and many analytical calculations can be done. From the practical point of view, it demands much less computer resources and is more suitable for an interface with external event generators \cite{kerbizi-lonnblad}. It is as rich as the model in Ref.\cite{kerbizi-2018}, depends on the same free parameters and, after retuning the latter, gives the same results.

The article is organized as follows. In Section \ref{sec:recursive-fragmentation} and \ref{sec:splitting matrix} the basis of the recursive polarized quark fragmentation model and the splitting matrix of the $string+{}^3P_0$ model are shortly described. The simplified version of the model is presented in Section \ref{sec:simple-3P0} and the comparison with Ref. \cite{kerbizi-2018} is discussed in Section \ref{sec:comparison-PRD}. In Section \ref{sec:positivity} the positivity conditions are analysed in the context of the present version of the model.

\section{Polarized recursive quark fragmentation}\label{sec:recursive-fragmentation}
The hadronization process $\QA \QbB\rightarrow h_1\dots h_N$ of the $\QA \QbB$ color neutral system can be thought to occur by the chain of splittings
\begin{eqnarray}\label{eq:qqbar-hadrons}
\nonumber \QA \rightarrow h_1 + q_2,..\,, q_r \rightarrow h_r + q_{r+1},..\,, q_{N-1}\rightarrow h_{N-1} +q_N,\\
\end{eqnarray}
$r$ being the ``rank'' of the hadron $h_r$. The chain terminates with $q_N+\QbB\rightarrow h_N$. The non-perturbative interaction between the initial quark and anti-quark is treated as a relativistic string with massless endpoints $\QA$ and $\QbB$. The decay of the string represents the hadronization of the $\QA\QbB$ system. In the center of mass frame of the $\QA\QbB$ system we orient the $\hat{\textbf{z}}$ axis along the momentum of $\QA$, which is also the jet or ``string'' axis.

The process in Eq. (\ref{eq:qqbar-hadrons}) is the recursive application of the elementary splitting
\begin{equation}
    \label{eq:splitting}
    q\rightarrow h + q'
\end{equation}
where $q$ is the current fragmenting quark, $h$ is the emitted hadron, with quark content $q\bar{q}'$, and $q'$ is the leftover quark. $h$ is restricted here to be a pseudoscalar meson. For a baryon $q'$ is replaced for instance by an anti-diquark. We denote by $k$ ($k'$) the four-momenta of $q$ ($q'$) and by $p$ the four-momentum of $h$. They are related by momentum conservation $k=p+k'$.

The process in Eq. (\ref{eq:splitting}) is described by the splitting function $F_{q',h,q}(Z,\pt;\kt,\textbf{S}_q)$ which gives the probability
\begin{equation}\label{eq:prob}
    dP_{q\rightarrow h+q'}=F_{q',h,q}(Z,\pt;\kt,\textbf{S}_q)\frac{dZ}{Z}d^2\pt
\end{equation}
that the hadron $h$ is emitted with forward light-cone momentum fraction $Z=p^+/k^+$ and with transverse momentum $\pt=\kt-\kpt$, and is normalized according to
\begin{eqnarray}\label{eq:normalization}
\sum_h\int_0^1\frac{dZ}{Z}\int d^2\pt F_{q',h,q}(Z,\pt;\kt,\textbf{S}_q)=1.
\end{eqnarray}
The light-cone momenta are defined as $p^{\pm}=p^0\pm p^3$. $\kt$ and $\kpt$ are the transverse momenta of $q$ and $q'$ with respect to the string axis. $p^-$ is not an independent variable but fixed by the mass-shell condition $p^-=\varepsilon_h^2/{p^+}$ where $\varepsilon_h^2= m_h^2+\ptpt$ is the hadrons transverse energy squared and $m_h$ is its mass.
We describe the quark spin states with Pauli spinors and encode the information on the quark polarization in the $2\times 2$ spin density matrix $\rho(q)=(1+\boldsymbol{\sigma}\cdot\textbf{S}_{q})/2$. The resulting ``polarized splitting function'' depends therefore on the polarization vector $\textbf{S}_q$. In Eq. (\ref{eq:prob}) the spin states of $q'$ are summed over.

The polarized splitting function can be calculated starting form the expression
\begin{equation}\label{eq:splitting-function}
    F_{q',h,q}= \rm{tr}\left[ T_{q',h,q}\, \rho(q)\, T^{\dagger}_{q',h,q}\right],
\end{equation}
where $T_{q',h,q}$ is a quantum mechanical ``splitting matrix'' acting on the quark $\rm{flavour} \otimes momentum \otimes spin$ space. Its elements are defined between the spin states of $q$ and of $q'$.
For practical applications, the splitting function in Eq. (\ref{eq:splitting-function}) is used for the generation of the hadron type $h$ and of its four-momentum, namely $Z$ and $\pt$, at the given momentum and polarization state of the quark $q$.
The spin density matrix of the leftover quark $q'$ is given by
\begin{equation}\label{eq:rho(q')}
    \rho(q')=\frac{T_{q',h,q}\, \rho(q)\, T^{\dagger}_{q',h,q}}{\rm{tr}\left[ T_{q',h,q}\, \rho(q)\, T^{\dagger}_{q',h,q}\right]}.
\end{equation}
The recursive application of Eq. (\ref{eq:splitting-function}) and of Eq. (\ref{eq:rho(q')}) in the Monte Carlo simulation allows to generate the hadron jets produced in the hadronization of polarized quarks \cite{kerbizi-2018}.

\section{Splitting matrix from the \\ general string + ${}^3P_0$ model}\label{sec:splitting matrix}
The string axis defines a privileged direction in space, thus the splitting matrix has not to be invariant under the full Lorentz group but only under the subgroup generated by rotations about the string axis (here $\hat{\textbf{z}}$), Lorentz boosts along the same axis and reflections about any plane containing it.

The splitting matrix, defined as \cite{kerbizi-2018}
\begin{eqnarray}\label{eq:T}
\nonumber T_{q',h,q}&=&C_{q',h,q}\,\check{g}(\varepsilon_h^2)\Delta_{q'}(\kpt)\Gamma_{h,s_h}\hat{u}_q^{-1/2}(\kt)\\
&\times&\left[(1-Z)/\varepsilon_h^2\right]^{a/2} \exp\left[-\bl\varepsilon_h^2/(2Z)\right],
\end{eqnarray}
respects these symmetries. 
The $Z$ dependence as required by LR symmetry is given in the second line and the parameters $a$ and $\bl$ are the same as in the LSM \cite{Lund1983}.

The factor $C_{q',h,q}$ describes the splitting of Eq. (\ref{eq:splitting}) in flavour space and is symmetric under the exchange of $q$ with $q'$, more precisely $C_{q',h,q}=C_{q,\bar{h},q'}$. It is proportional to the meson wave function $\langle q\bar{q}'|h\rangle$ in flavour space and also takes into account the suppression of strange mesons and the suppression of $\eta$ with respect to $\pi^0$.

The complex $2\times 2$ matrix in quark spin space
\begin{eqnarray}\label{eq:delta}
\Delta_q(\kt)=(\mu_q+\sigma_z\boldsymbol{\sigma}\cdot  \kt) f_{\T}(\ktkt)
\end{eqnarray}
gives the $\kt$-dependent part of the quark propagator inspired to the $^3P_0$ mechanism. It depends on the complex mass parameter $\mu_q$ which is responsible for the single spin effects. We take the same complex parameter for all quark flavours, i.e. $\mu_q\equiv \mu$.
The function $f_{\T}$ is a fast decreasing function of the quark transverse momentum at the string breaking. In Ref. \cite{kerbizi-2018} it has been taken as
\begin{equation}\label{eq:fT_gauss}
    f\T(\ktkt)=\sqrt{\frac{\bt}{\pi}}\exp(-\bt\ktkt/2).
\end{equation}
It depends only on the parameter $\bt$ which is related to the width of the quark (and anti-quark) transverse momentum at each string breaking. This choice of $f\T$ leads to an exponential decay of the hadrons $p^2\T$ spectrum.
The same function was proposed in Ref. \cite{Field-Feynman} but other choices are possible. For instance in the SLM it comes out to be a correlated gaussian in the transverse momenta of two successive quarks \cite{Andersson-kt-correlations} while in the event generator PYTHIA the quark $\kt$ at string breaking is generated according to the function $p_0\exp(-\ktkt/\sigma_0^2)+p_1\exp(-\ktkt/\sigma_1^2)$.
An alternative class of functions is
\begin{equation}\label{eq:fT_alpha}
    f_{\rm{T}}(\ktkt)\propto \frac{\exp(-\bt\ktkt/2)}{\big( |\mu|^2+\ktkt\big)^{\alpha}},
\end{equation}
where the denominator is inspired from the Feynman propagator $1/(\gamma\cdot k-m_q)=(\gamma\cdot k + m_q)/(k^2-m_q^2)$, the analog of $(k^2-m_q^2)$ being $-(\ktkt+|\mu|^2$). This analogy suggests $\alpha=1$, but in principle any power is allowed: $\alpha=0$ brings back to Eq.(\ref{eq:fT_gauss}),
$\alpha\neq 0$ modifies the tail in the $p^2\T$ distribution of the hadrons. We have performed simulations using both Eq. (\ref{eq:fT_gauss}) and Eq. (\ref{eq:fT_alpha}) for different values of $\alpha$ obtaining predictions only slightly different, allowing the choice $\alpha=0$ of Ref. \cite{kerbizi-2018}.

The matrix $\Gamma_{h,s_h}$ is the vertex matrix which describes the $q-h-q'$ coupling. It depends on the hadron spin state $s_h$ and possibly on $\kt$ and $\kpt$, at most as a polynomial. Neglecting the latter possibility, the coupling for pseudo-scalar meson emission is
\begin{equation}\label{eq:pseudo-scalar coupling}
    \Gamma_h=\sigma_z,
\end{equation}
analogous to the Dirac $\gamma_5$ coupling.

The matrix $\hat{u}_q(\kt)$ is related to the single quark density in momentum $\otimes$ spin space and can be written as \cite{kerbizi-2018}
\begin{eqnarray}\label{eq:uhat}
\nonumber \hat{u}_q(\kt)&=&\sum_{h\,}|C_{q',h,q}|^2\int d^2\kpt \check{g}^2(\varepsilon_h^2) N_a(\varepsilon_h^2)\\
&\times&\sum_{s_h} \Gamma_{h,s_h}^{\dagger} \Delta_{q'}(\kpt)^{\dagger} \Delta_{q'}(\kpt) \Gamma_{h,s_h}\\
&\equiv& \hat{u}_{0q}(\ktkt)+\hat{u}_{1q}(\ktkt)\boldsymbol{\sigma}\cdot\ntil,
\end{eqnarray}
where $\ntil(\kt)=\hat{\textbf{z}}\times \kt/|\kt|$ and
\begin{eqnarray}\label{eq:Na}
N_a(\varepsilon_h^2)=\int_0^1 dZZ^{-1} \left(\frac{1-Z}{\varepsilon_h^2}\right)^{a} \exp\left(-\bl\frac{\varepsilon_h^2}{Z}\right).
\end{eqnarray}
The matrix $\hat{u}_q$ is positive definite, with $\hat{u}_{0q}>|\hat{u}_{1q}|$, and allows the splitting function to be normalized according to Eq. (\ref{eq:normalization}). The insertion of $\hat{u}_q^{-1/2}$ in Eq. (\ref{eq:T}) is necessary to fulfill the LR symmetry requirement.

The model allows for different choices of the function $\check{g}(\varepsilon_h^2)$.
For a general form of $\check{g}$,
\begin{itemize}
\item[i.] there are dynamical spin-independent $\kt$-$\kpt$ correlations \cite{artru-belghobsi-essma}
\item[ii.] the generation of the hadron type depends on $\textbf{S}\T$ and $\kt$ 
\end{itemize}
In Ref. \cite{kerbizi-2018} we choosed $\check{g}(\varepsilon_h^2)=(\varepsilon_h^2)^{a/2}$. It gives properties (i) and (ii). To simplify, in the Monte Carlo implementation of Ref. \cite{kerbizi-2018} the point (ii) was not considered, introducing some breaking of the LR symmetry. This is the free input function that has been revised in the present work leading to a simplification of the formalism.

\section{The simplified string + ${}^3P_0$ model}\label{sec:simple-3P0}
In the present work we choose a different $\check{g}$-function, namely
\begin{equation}
    \label{eq:choice-g}
    \check{g}(\varepsilon_h^2)=1/\sqrt{N_a(\varepsilon_h^2)}
\end{equation}
which was already quoted as a possible choice in Ref.\cite{kerbizi-2018}.

With the present definition of $\check{g}$ the generation of the hadron type does not depend on the fragmenting quark transverse momentum and on its transverse polarization and there are no spin-independent dynamical correlations between the transverse momenta of two successive quarks, as in the SLM and in the model of Ref. \cite{DS09}. However we gain in simplicity while satisfying exactly the LR symmetry.

Note that the property (i) can be re-introduced by taking $\check{g}(\varepsilon_h^2)=e^{-b_1\varepsilon_h^2}/\sqrt{N_a(\varepsilon_h^2)}$, where $b_1$ is a new parameter describing the spin independent correlations. In this case, the $Z$-integrated $\pt$ distribution of the splitting function remains simple and, taking the same $\mu_q$ for all flavors, only the relative probability (vector meson) /(pseudo-scalar meson) depends on $\kt$ and on $\textbf{S}_{q\rm{T}}$ of the parent quark. Such a way of introducing $\kt-\kpt$ correlations is also used in Ref. \cite{Andersson-kt-correlations} in the spinless SLM model. However, presently there is no compelling reason to introduce the spin-independent $\kt-\kpt$ correlations.

Our choice for $\check{g}$ is also in line with the implementation of the LSM in $\rm{PYTHIA}$ \cite{pythia8}. Thus it is more suitable in view of the inclusion of spin effects in the hadronization of this event generator \cite{kerbizi-lonnblad}. Also, it allows for a simpler description of the spin transfer mechanism, as will be shown in the following.

Equation (\ref{eq:choice-g}) introduces a remarkable simplification with respect to Ref. \cite{kerbizi-2018}, in particular the matrix $\hat{u}_q$ of Eq. (\ref{eq:uhat}) becomes proportional to the unit matrix. With only pseudo-scalar mesons and Eq. (\ref{eq:pseudo-scalar coupling}), it is
\begin{eqnarray}\label{eq:hat-u-explicit}
\hat{u}_q(\kt)= \textbf{1}\, \sum_h |C_{q',h,q}|^2 \,\langle |\mu|^2+\kptkpt\rangle_{\T},
\end{eqnarray}
where we have defined the average operation
\begin{equation}
\langle g\rangle_{\T}=\int d^2 \kt g(\ktkt) f^2_{\T}(\ktkt)
\end{equation}
for a generic function $g$.

Using Eqs. (\ref{eq:splitting-function},\ref{eq:T}-\ref{eq:fT_gauss}), the splitting function becomes
\begin{eqnarray}\label{eq:F_explicit}
 \nonumber F_{q',h,q}(Z,\pt;\kt,\textbf{S}_q)&=& \frac{|C_{q',h,q}|^2}{\sum_H |C_{q',H,q}|^2}  \\
 &\times& \frac{|\mu|^2+\kptkpt}{\langle |\mu|^2+\kptkpt\rangle_{\T}} f_{\T}^2(\kptkpt)\\
\nonumber &\times&\left[ 1-\frac{2\IM(\mu)\,\rm{k'}_{\rm{T}}}{|\mu|^2+\kptkpt}\textbf{S}_{q}\cdot\tilde{\textbf{n}}(\kpt) \right]\\
\nonumber &\times& \left(\frac{1-Z}{\varepsilon_h^2}\right)^a \frac{\exp{(-\bl \varepsilon_h^2/Z)}}{N_a(\varepsilon_h^2)},
\end{eqnarray}
where the third line is source of the Collins effect in the model.
The splitting function satisfies the normalization condition in Eq. (\ref{eq:normalization}) and is much simpler than the one given by Eqs. (52-54) of Ref. \cite{kerbizi-2018}.

In this new version of the model it is more convenient to draw the hadron $h$ generating first its type according to the first line of Eq. (\ref{eq:F_explicit}), then the transverse momentum $\pt=\kt-\kpt$ according to the second and third lines and then finally the longitudinal momentum fraction $Z$ according to the last line of Eq. (\ref{eq:F_explicit}). In Ref. \cite{kerbizi-2018} the simplest order was the hadron type first, then $Z$ and finally $\pt$.

As already mentioned, with the choice of Eq. (\ref{eq:choice-g}), there is no spin-independent correlation between $\kt$ and $\kpt$ in the $Z$-integrated splitting function. The only source of correlation between $\kt$ and $\kpt$ comes from the ${}^3P_0$ mechanism associated to the correlation between the spins of $q$ and $\bar{q}'$ in the hadron. For a pseudo-scalar hadron it gives $\langle \kt\cdot\kpt\rangle<0$, i.e. on the average $\kt$ and $\kpt$ are anti-parallel.


The polarization vector of the leftover quark $q'$ can then be calculated from Eq. (\ref{eq:rho(q')}). The transverse and the longitudinal components are
\begin{eqnarray}\label{eq:S_q'T}
\nonumber \textbf{S}_{q'\rm{T}}=\frac{1}{N}\big[&-&(|\mu|^2+\kptkpt)\,\textbf{S}_{q\rm{T}}+2(\textbf{S}_{q\rm{T}}\cdot\,\kpt)\kpt\\
&-&2\IM(\mu)\,\rm{k'_T}\,\tilde{\textbf{n}}(\kpt)-2\RE{\mu}\,S_{qz}\,\kpt\big],
\end{eqnarray}
\begin{eqnarray}\label{eq:S_q'L}
S_{q'z}=\frac{1}{N}\big[(|\mu|^2-\kptkpt)\,S_{qz}-2\RE{\mu}\,\textbf{S}_{q\rm{T}}\cdot\kpt]
\end{eqnarray}
where the normalization $N$ is given by
\begin{equation}\label{eq:N}
    N=|\mu|^2+\kptkpt-2\IM{\mu}\,\rm{k'_T}\,\textbf{S}_{q\rm{T}}\cdot \tilde{\textbf{n}}(\kpt).
\end{equation}
From Eq. (\ref{eq:S_q'T}) it is clear that the transverse polarization of $q'$ has several different types of contributions: it inherits some (depending on $\kpt$) of the transverse polarization of $q$ but can also receive contributions from $\kpt$ alone. In addition, there can be a transfer from longitudinal to transverse polarization and vice-versa. If the quark $q$ is in a pure state $(\textbf{S}_q^2=1)$, then also $q'$ will be in a pure state ($\textbf{S}_{q'}^2=1$). This is due to the fact that the emitted meson has spin zero, thus cannot take spin information away. 

If the transverse momentum of $q'$ is integrated over there is a leakage of spin information on $q'$ ($\kpt$ is correlated with $\textbf{S}_q'$) and the quark polarization decays along the fragmentation chain.
Therefore, at each step of the recursive process both the quark transverse and longitudinal polarizations decay. 

The polarized decay process is described by the transverse and the longitudinal depolarization factors $D_{\rm{TT}}$ and $D_{\rm{LL}}$. They are obtained from Eqs. (\ref{eq:S_q'T})-(\ref{eq:S_q'L}) integrating over $\kpt$ separately the numerator and the denominator. The analytic expressions are
\begin{eqnarray}\label{eq:DTT}
\textbf{S}_{q'\rm{T}}&=&-\frac{\langle |\mu|^2\rangle_{\rm{T}}}{\langle |\mu|^2+\kptkpt\rangle_{\rm{T}}}\textbf{S}_{q\rm{T}}\equiv D_{\rm{TT}}\,\,\textbf{S}_{q\rm{T}}
\end{eqnarray}
\begin{eqnarray}\label{eq:DLL}
S_{q'z}=\frac{\langle |\mu|^2-\kptkpt\rangle\T}{\langle |\mu|^2+\kptkpt\rangle\T}\,S_{qz}\equiv D_{\rm{LL}}\,S_{qz}.
\end{eqnarray}
The depolarization factors depend on the complex mass and on the width of quark transverse momentum $\kptkpt$, ie. on the choice of the function $f\T$. For $f\T$ of Eq. (\ref{eq:fT_gauss}) it is $D_{TT}=-\bt|\mu|^2/(\bt|\mu|^2+1)$ and $D_{LL}=(\bt|\mu|^2-1)/(\bt|\mu|^2+1)$ as in Ref. \cite{DS09}.
We note that $D_{\rm{TT}}<0$ as expected for the production of a pseudo-scalar meson in the string+${}^3P_0$ model. This gives Collins effects of opposite sign for even and odd rank mesons.

\section{Comparison with the previous results}\label{sec:comparison-PRD}
As for the previous model \cite{kerbizi-2018}, we have implemented the present choice of the function $\check{g}$ in a recursive stand alone Monte Carlo. The code is the same except for the routines used for the generation of $Z$ and $\kpt$ which have been changed according to Eq. (\ref{eq:F_explicit}). The free parameters are the same and have the same values as in Ref. \cite{kerbizi-2018} except for $\bt$. In particular $a=0.9$, $\bl=0.5\,\rm{GeV}^{-2}$, $\mu=(0.42+i\,0.76)\,\rm{GeV}$ and $\bt=8.43\,\rm{GeV}^{-2}$ which is $1.63$ times larger than the value used in Ref. \cite{kerbizi-2018} in order to have similar $p_T^2$ distributions in spite of the different choices for $\check{g}$. The increase in $\bt$ is necessary to compensate the exponential growth, at large $\ptpt$, of $\check{g}(\varepsilon_h^2)$ given by Eq. (\ref{eq:choice-g}).  

The results shown in the next sections are obtained from simulations of the fragmentation of fully transversely polarized $u$ quarks whose momentum is determined using the same sample of $x_B$ and $Q^2$ values of SIDIS events as in Ref. \cite{kerbizi-2018}.

\subsection{Kinematical distributions}
The rank dependence of the kinematical distributions comes from the recursive nature of the model and is about the same as that in Ref. \cite{kerbizi-2018}. In particular, the $Z$ and $\ptpt$ distributions do not depend on the rank for $r\geq 2$.

\begin{figure*}
    \centering
    \begin{subfigure}[b]{0.35\textwidth}
        \includegraphics[width=\textwidth]{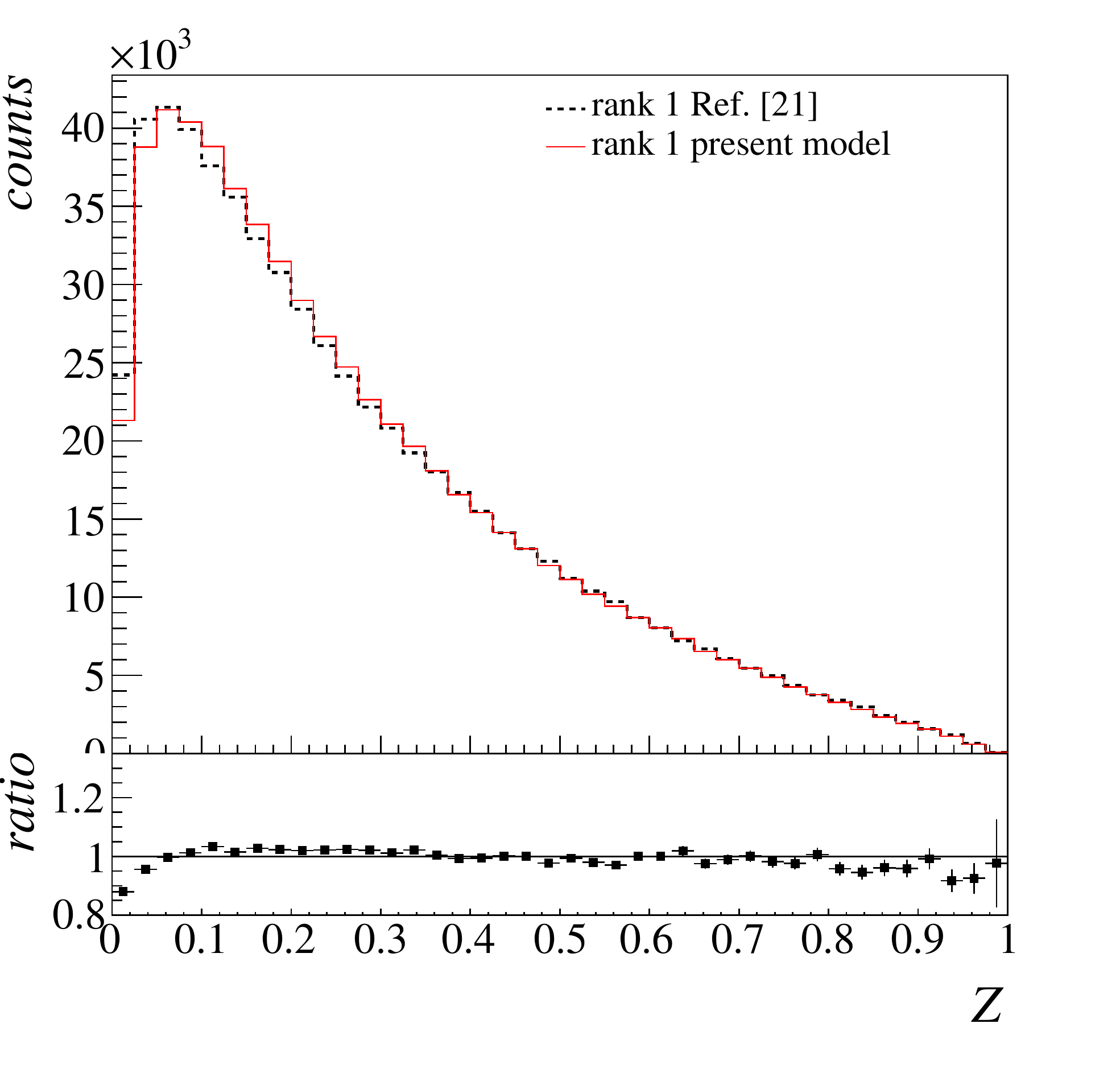}
       \caption{}
    \end{subfigure}
     \begin{subfigure}[b]{0.35\textwidth}
        \includegraphics[width=\textwidth]{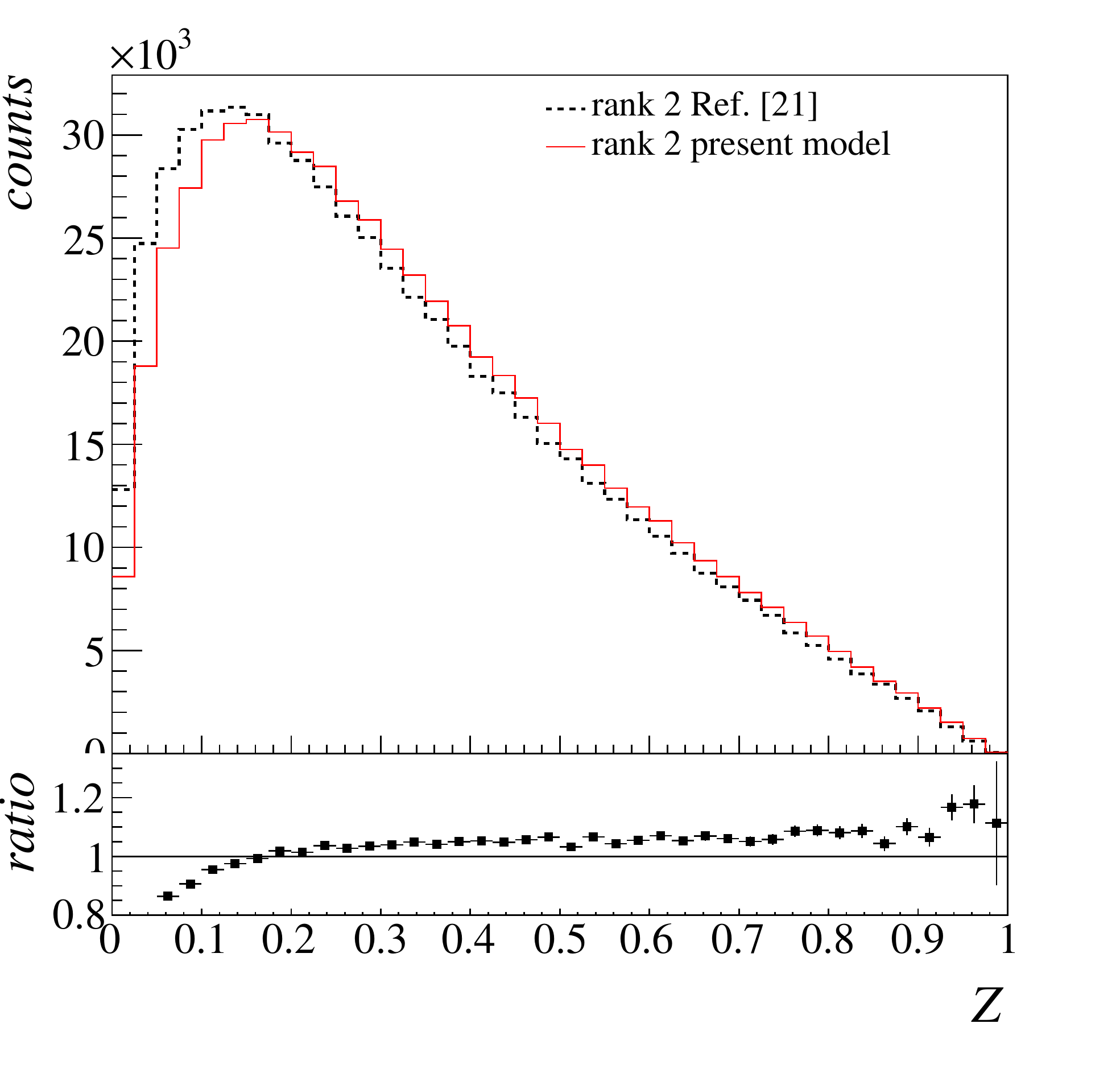}
        \caption{}
    \end{subfigure}
    \begin{subfigure}[b]{0.35\textwidth}
        \includegraphics[width=\textwidth]{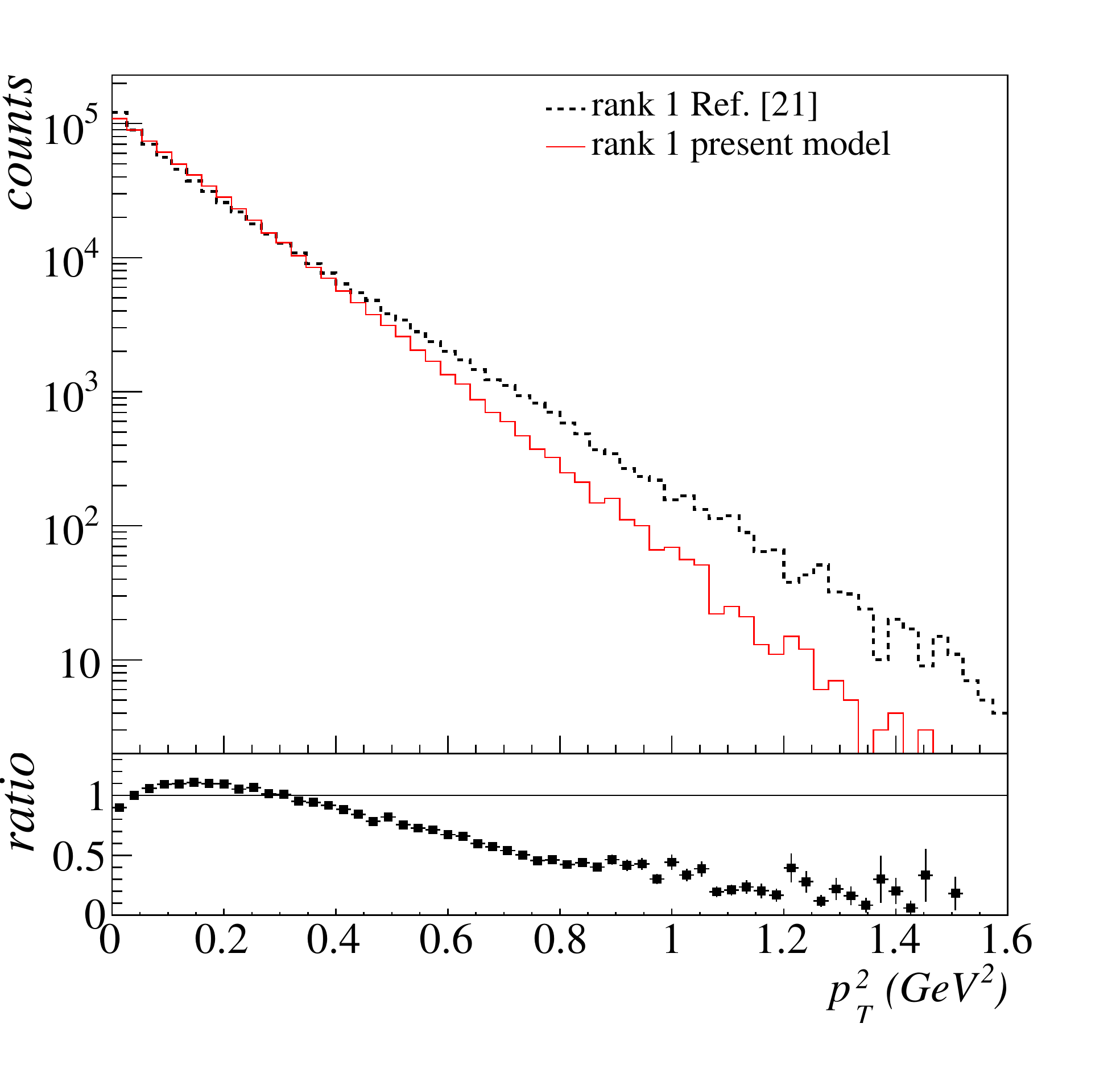}
       \caption{}
    \end{subfigure}
    \begin{subfigure}[b]{0.35\textwidth}
        \includegraphics[width=\textwidth]{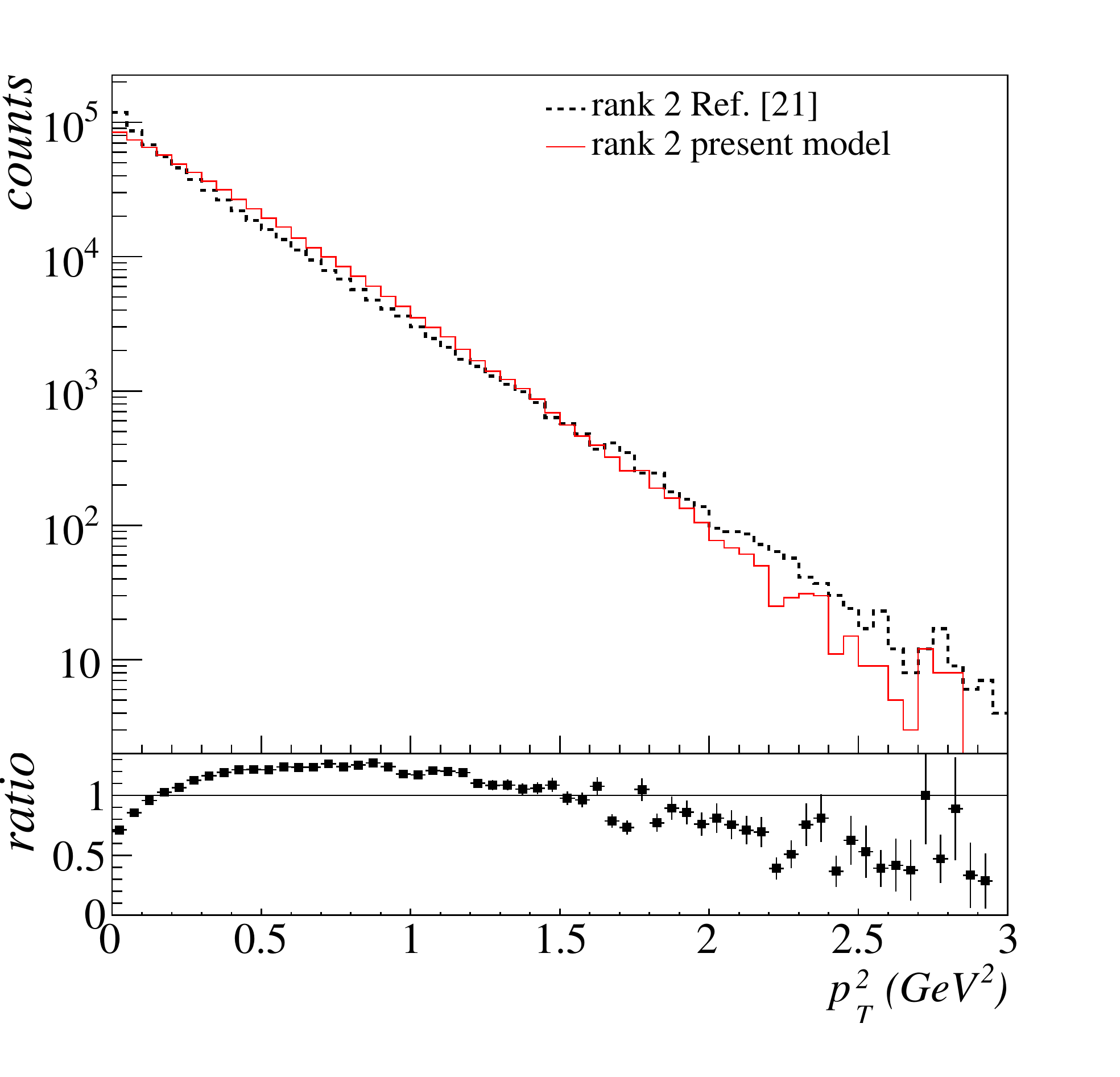}
       \caption{}
    \end{subfigure}
    
 \caption{Comparison between the model of Ref. \cite{kerbizi-2018} (dotted histogram) and the simplified ${}^3P_0$ (continous histogram) for: (a) $Z$ distribution for rank 1 hadrons, (b) $Z$ distribution for rank 2 hadrons, (c) $p_T^2$ distribution for rank 1 hadrons and (d) $p_T^2$ distribution for rank 2 hadrons. Their ratios are shown in the bottom plots. Note the different horizontal scales in plots (c) and (d).}\label{fig:Z_pT2_rank}   
\end{figure*}

In Fig. \ref{fig:Z_pT2_rank} we compare the $Z$ and $\ptpt$ distributions for the $r=1$ (left plots) and $r=2$ (right plots) hadrons as obtained with the present model (continuous histograms) and with the model of Ref. \cite{kerbizi-2018} (dotted histograms). Their ratio is shown in the bottom plot of each panel. The two models produce almost the same $Z$ distribution for rank 1 (plot (a)) as expected because the initial quark does not have $\kt$. For rank 2 (plot (b)) the $Z$ distribution in this model is slightly shifted towards greater values of $Z$. This is correlated to a somewhat larger $\langle \ptpt \rangle$, as can be seen from plot (d).

From plot (c) it is also clear that the $\ptpt$ distribution for rank $1$ of Ref. \cite{kerbizi-2018} has two slopes on the contrary to this model. In fact the $\ptpt$ distribution of Ref. \cite{kerbizi-2018} is a sum of contributions of different slopes, one for each $Z$, due to the factor $\exp(-\bl\varepsilon_h^2/Z)$. In the present model also there is a different slope for each $Z$, but the factor $1/N_a(\varepsilon_h^2)$ ``rectifies'' the slope of the $Z$-integrated $\ptpt$ spectrum. 

\begin{figure*}
    \centering
    \begin{subfigure}[b]{0.35\textwidth}
        \includegraphics[width=\textwidth]{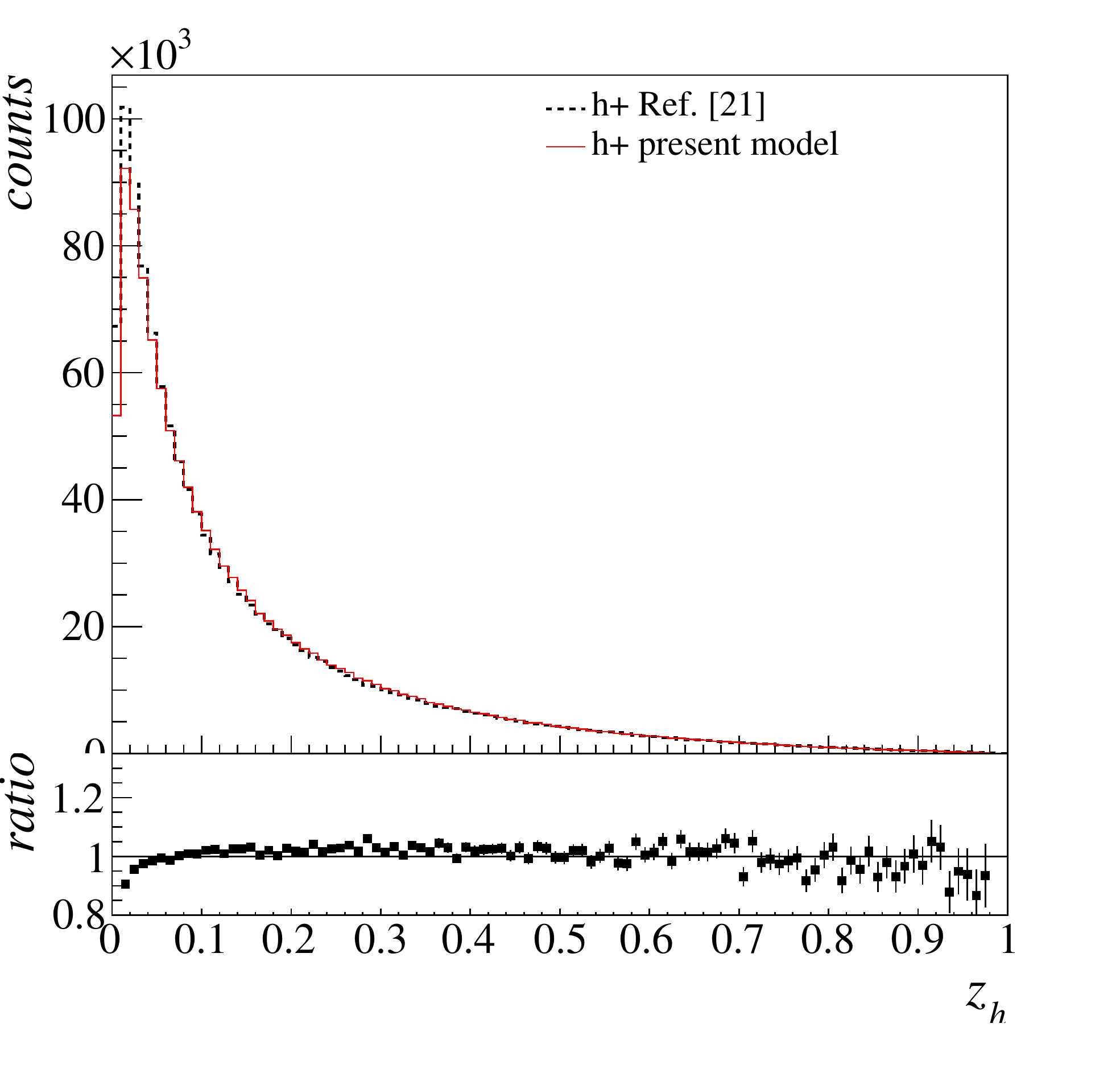}
    \end{subfigure}
    \begin{subfigure}[b]{0.35\textwidth}
        \includegraphics[width=\textwidth]{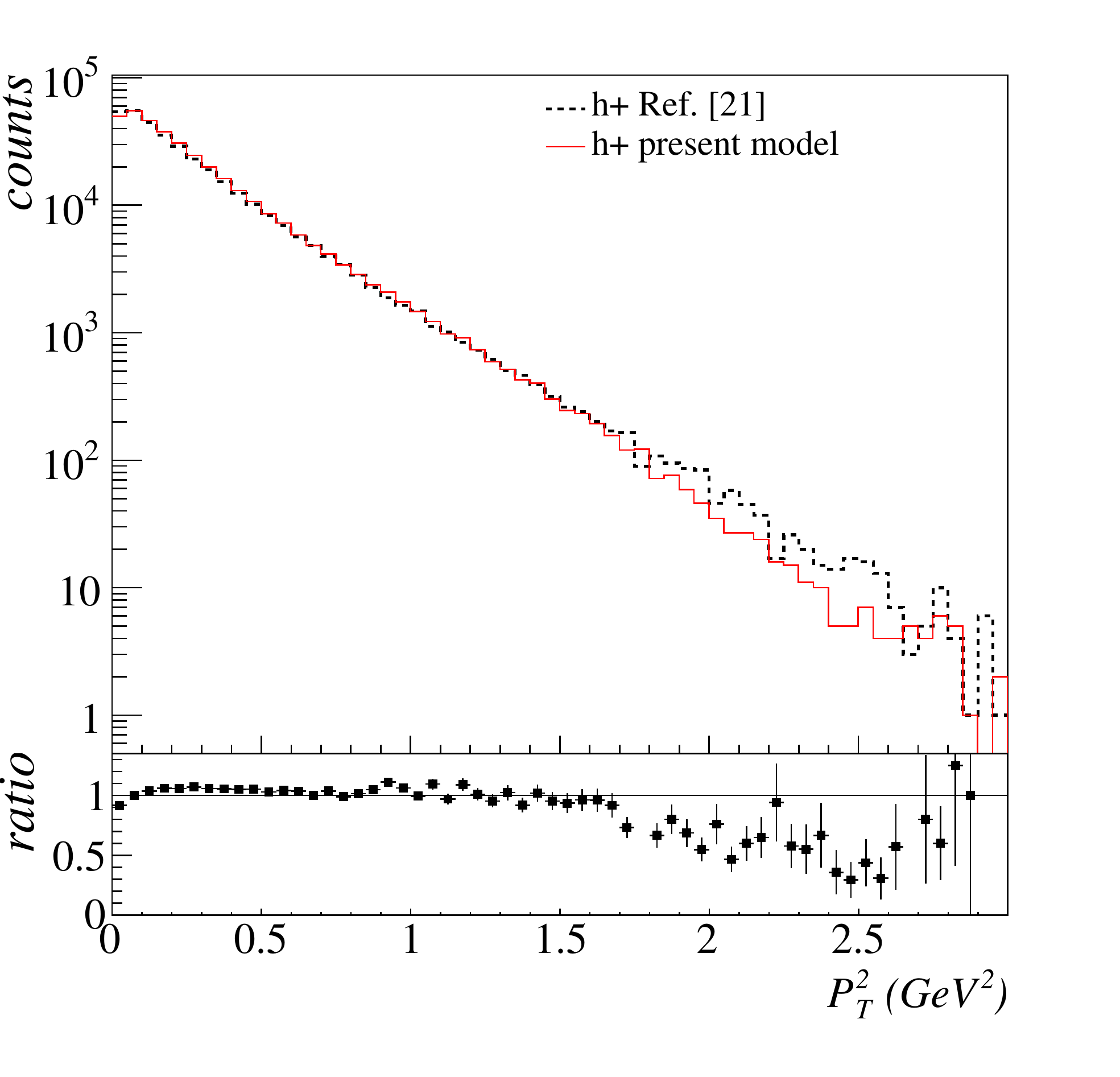}
    \end{subfigure}

 \caption{Comparison between $z_h$ (left plot) and $p_T^2$ (right plot) distributions of positively charged hadrons as obtained with the model of Ref. \cite{kerbizi-2018} (dotted histogram) and with the simplified ${}^3P_0$ (continous histogram). Their ratio is also displayed in the respective bottom panels. We have applied the cuts $z_h>0.2$ and $p\T>0.1\,GeV$.}\label{fig:zh_pT2_hadrons}   
\end{figure*}

The differences are even smaller when looking at measurable quantities. The distributions of the fraction $z_h$ of the fragmenting quark energy taken by the positive hadron in the two models are shown in the left plot of Fig. \ref{fig:zh_pT2_hadrons}. The region of very small $z_h$ is less populated in the simplified ${}^3P_0$ model.
The $\ptpt$ distribution for positive hadrons is almost the same in both models as shown in the right plot of Fig. \ref{fig:zh_pT2_hadrons}.

\begin{figure}[htb]
  \centering
    \includegraphics[width=0.35\textwidth]{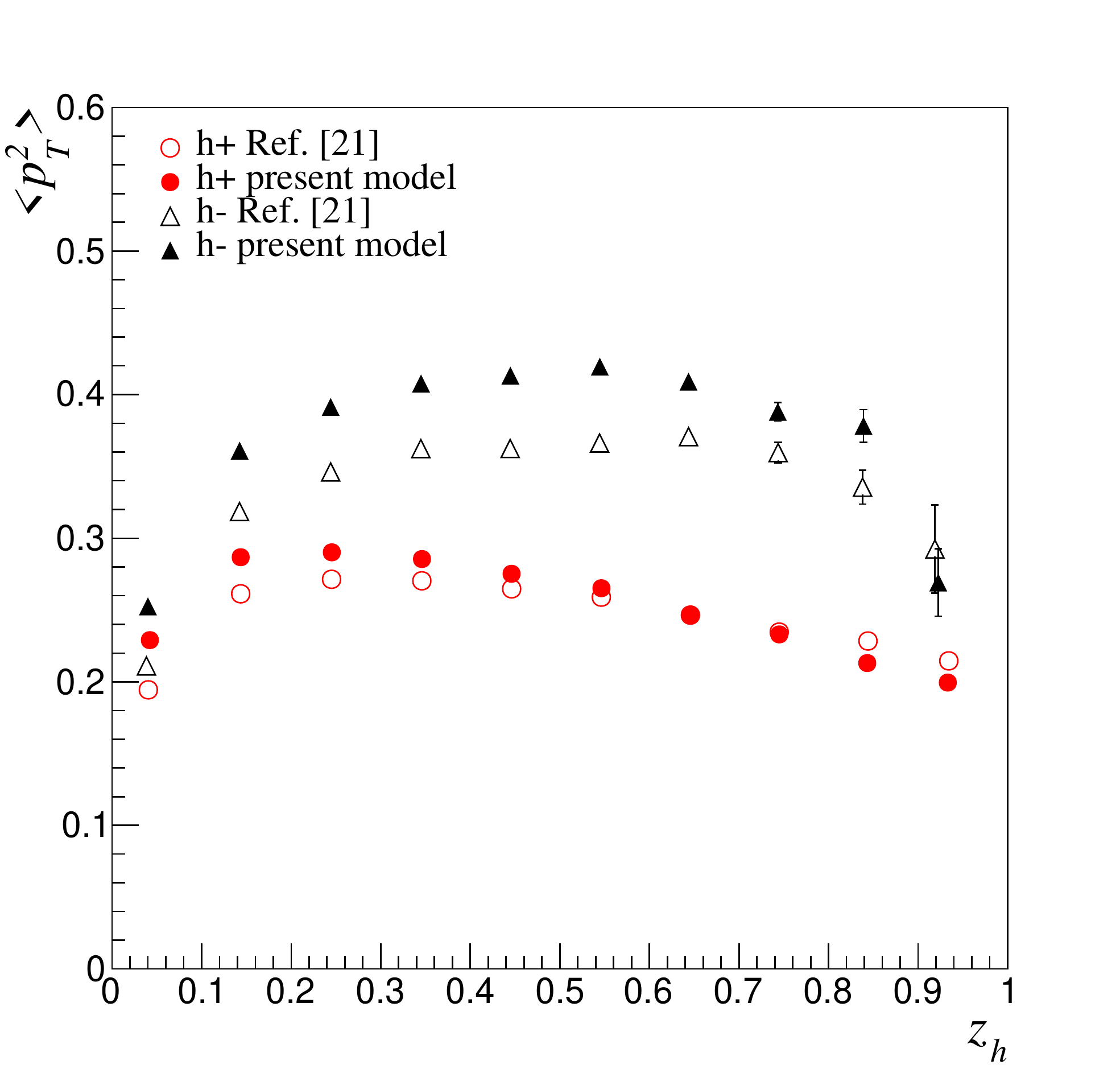}
  \caption{Comparison between the $z_h$ dependence of $\langle p_T^2\rangle$ in the present model (full points) and in the model of Ref. \cite{kerbizi-2018} (open points).}\label{fig:pt2_zh}
\end{figure}

Figure \ref{fig:pt2_zh} compares the $z_h$ dependence of the transverse momentum width $\langle p_T^2\rangle$ of charged hadrons in the two models. The present model gives a larger difference between the $\langle p_T^2\rangle$ for positive hadrons and the $\langle p_T^2\rangle$ for negative hadrons than the model of Ref. \cite{kerbizi-2018}, which already was not in agreement with experiments. Indeed, due to the pure spin correlations it is $\langle \kt\cdot\kpt\rangle <0$, now at ranks larger than one we have $\langle \ptpt \rangle > 2\langle \ktkt\rangle$. In Ref. \cite{kerbizi-2018}, on the other hand, the spin-independent correlation, if taken alone, would give the opposite correlation $\langle \kt\cdot\kpt\rangle >0$, therefore $\langle \ptpt \rangle<2\langle \ktkt\rangle$.

\subsection{Single hadron transverse spin asymmetries}
Hadrons in the fragmentation of transversely polarized quarks exhibit a left-right asymmetry with respect to the plane defined by the transverse spin and the momentum of the quark, according to the azimuthal distribution
\begin{eqnarray}\label{eq:collins_distribution}
\frac{dN_h}{dz_hd^2\pt}\propto 1+a^{q_A\uparrow\rightarrow h+X}S_{\rm{AT}}\sin\phi_C
\end{eqnarray}
where $a^{q_A\uparrow\rightarrow h+X}$ is the Collins analysing power for hadron $h$, $S_{\rm{AT}}$ is the transverse polarization of the fragmenting quark $q_A$ and $\phi_C=\phi_h-\phi_{S_{\rm{AT}}}$ is the Collins azimuthal angle. Being formulated at the amplitude level, this model produces a pure $\sin\phi_C$ modulation.

\begin{figure}[tb]
  \centering
    \includegraphics[width=0.4\textwidth]{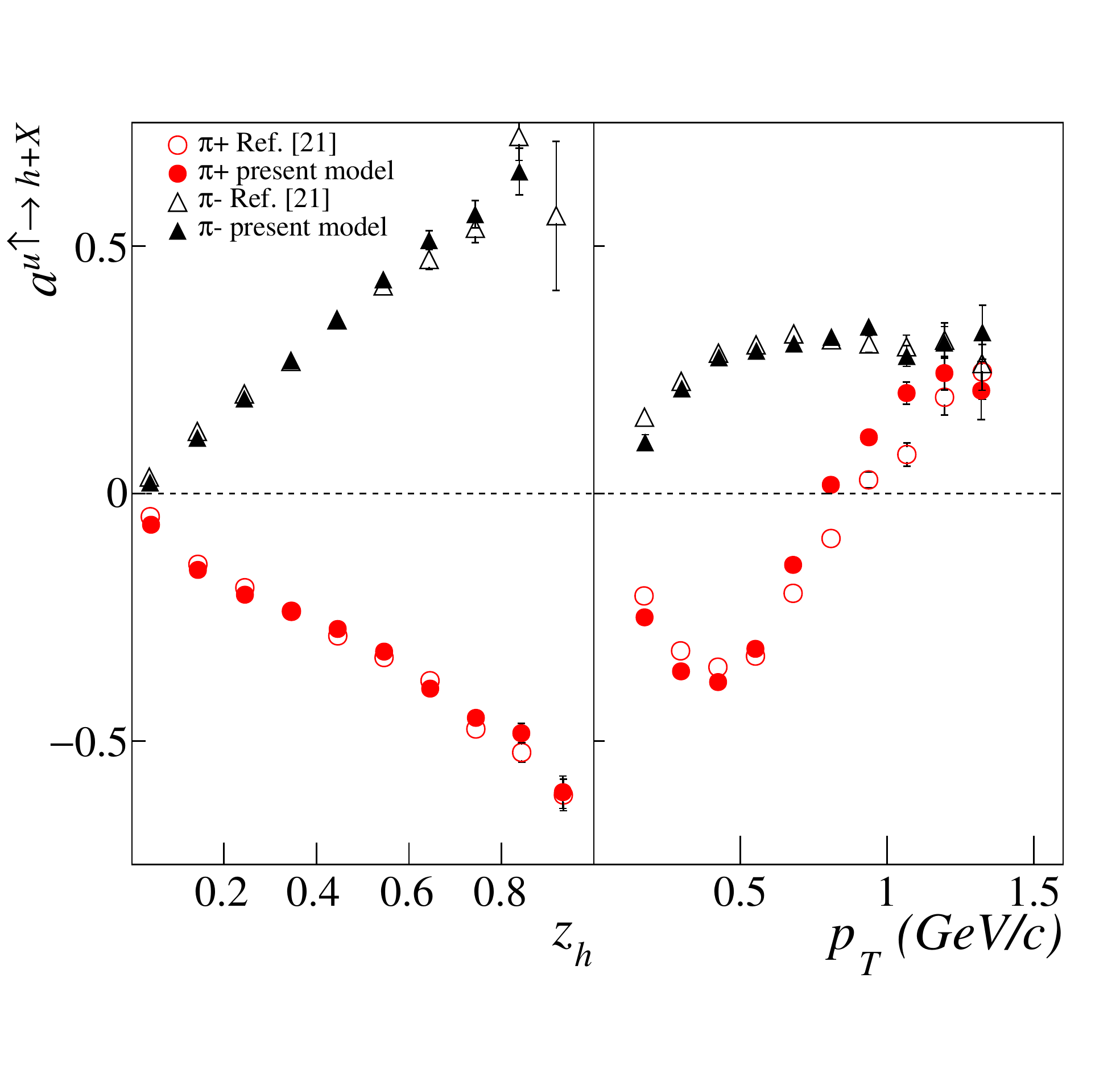}
  \caption{Collins analysing power for charged pions as function of $z_h$ (left panel) and $p_T$ (right panel) as obtained with the present model (full points) and with the model of Ref. \cite{kerbizi-2018} (open points). The cuts $z_h>0.2$ and $p_T>0.1\, GeV$ have been applied.}\label{fig:collins_asymmetry}
\end{figure}

Figure \ref{fig:collins_asymmetry} shows the Collins analysing power for charged pions produced in jets of transversely polarized $u$ quarks estimated as $2\langle \sin\phi_C\rangle$ (full points). They are compared with the results of Ref. \cite{kerbizi-2018} (open points). The analysing power is shown as function of $z_h$ in the left plot and as function of $p_T$ in the right plot of Fig \ref{fig:collins_asymmetry}. The cuts $z_h>0.2$ and $p_{\rm{T}}>0.1\, GeV$ have been applied. Both models produce the same features for the analysing power. Some slight differences can be seen for the analysing power as function of $p_T$ for $\pi^+$ (right plot) which are due to the different ${k'}^2_T$ dependencies of the respective splitting functions.

\begin{figure}[tb]
  \centering
    \includegraphics[width=0.35\textwidth]{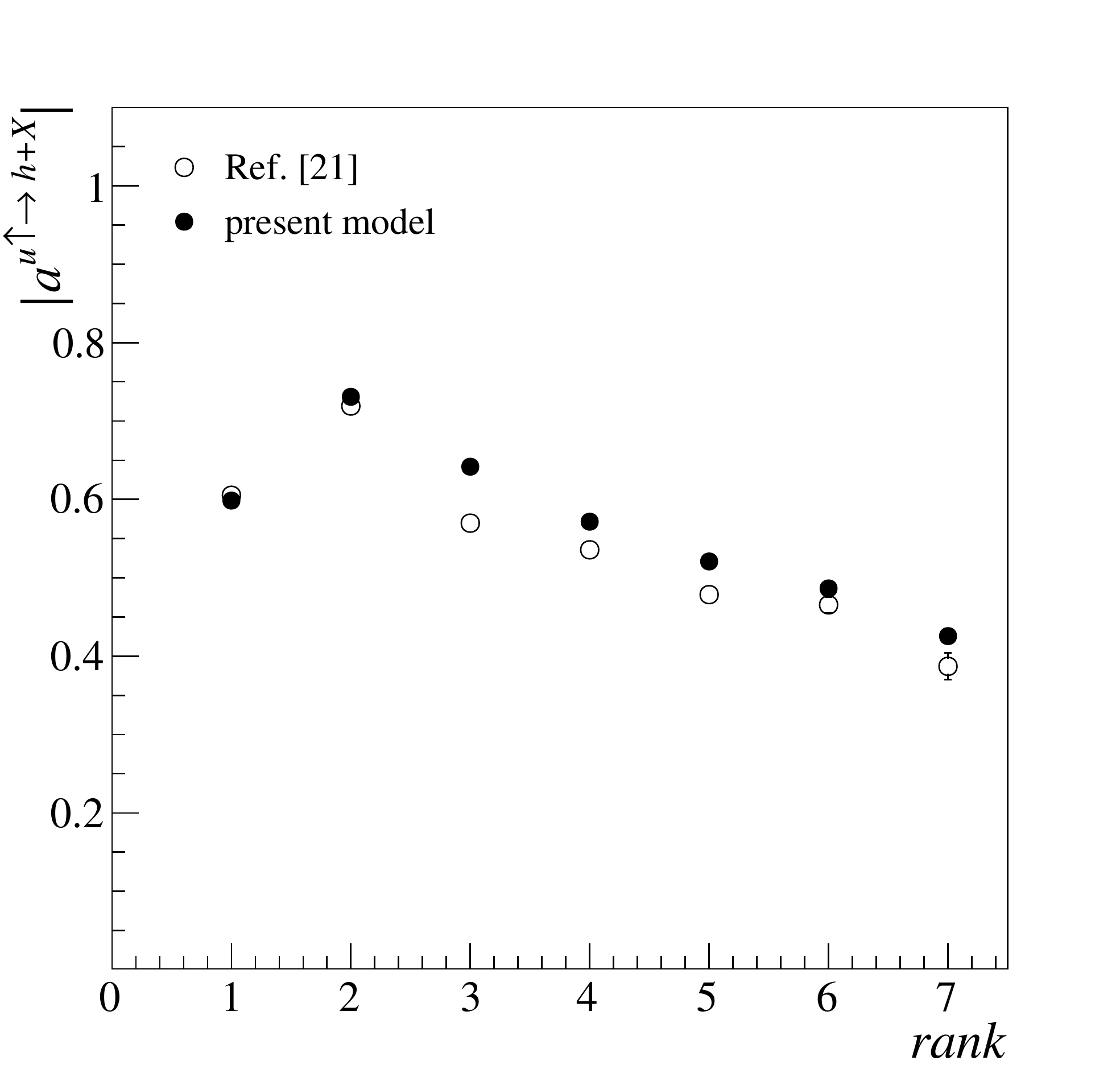}
  \caption{Comparison of the absolute value of the Collins analysing power as function of rank as obtained with the present model (full points) and with the model of Ref. \cite{kerbizi-2018} (open points). The cuts $z_h>0.2$ and $p_T>0.1\,GeV$ have been applied.}\label{fig:collins_asymmetry_rank}
\end{figure}

The absolute value of the Collins analysing power as function of the rank is shown in Fig. \ref{fig:collins_asymmetry_rank} for the present model (full points) and for the model of Ref. \cite{kerbizi-2018} (open points). In the present model the analysing power decays slower because of the triviality of the $\hat{u}_q$ matrix.


\subsection{Di-hadron transverse spin asymmetry}
The azimuthal distribution of hadron pairs of opposite charge in the same jet produced in the fragmentation of a transversely polarized quark is described by the equation
\begin{eqnarray}
\nonumber \frac{dN_{h_1h_2}}{dz\,dM_{inv}d\phi_R}\propto 1+a^{q\A\uparrow \rightarrow h_1h_2+X}\,S_{\rm{AT}}\sin(\phi_R-\phi_{\textbf{S}_{\rm{AT}}})\\
\end{eqnarray}
where $z=z_{h_1}+z_{h_2}$ is the sum of the fractional energies of the positive ($h_1$) and negative ($h_2$) hadrons and $M_{inv}$ is the invariant mass of the pair. The angle $\phi_R$ is the azimuthal angle of the transverse vector $\textbf{R}\T=(z_{h_2}\textbf{p}_{1\rm{T}}-z_{h_1}\textbf{p}_{2\rm{T}})/z$. $\textbf{p}_{1\rm{T}} (\textbf{p}_{2\rm{T}})$ is the transverse momentum of the positively (negatively) charged hadron of the pair.

\begin{figure}[tb]
  \centering
    \includegraphics[width=0.5\textwidth]{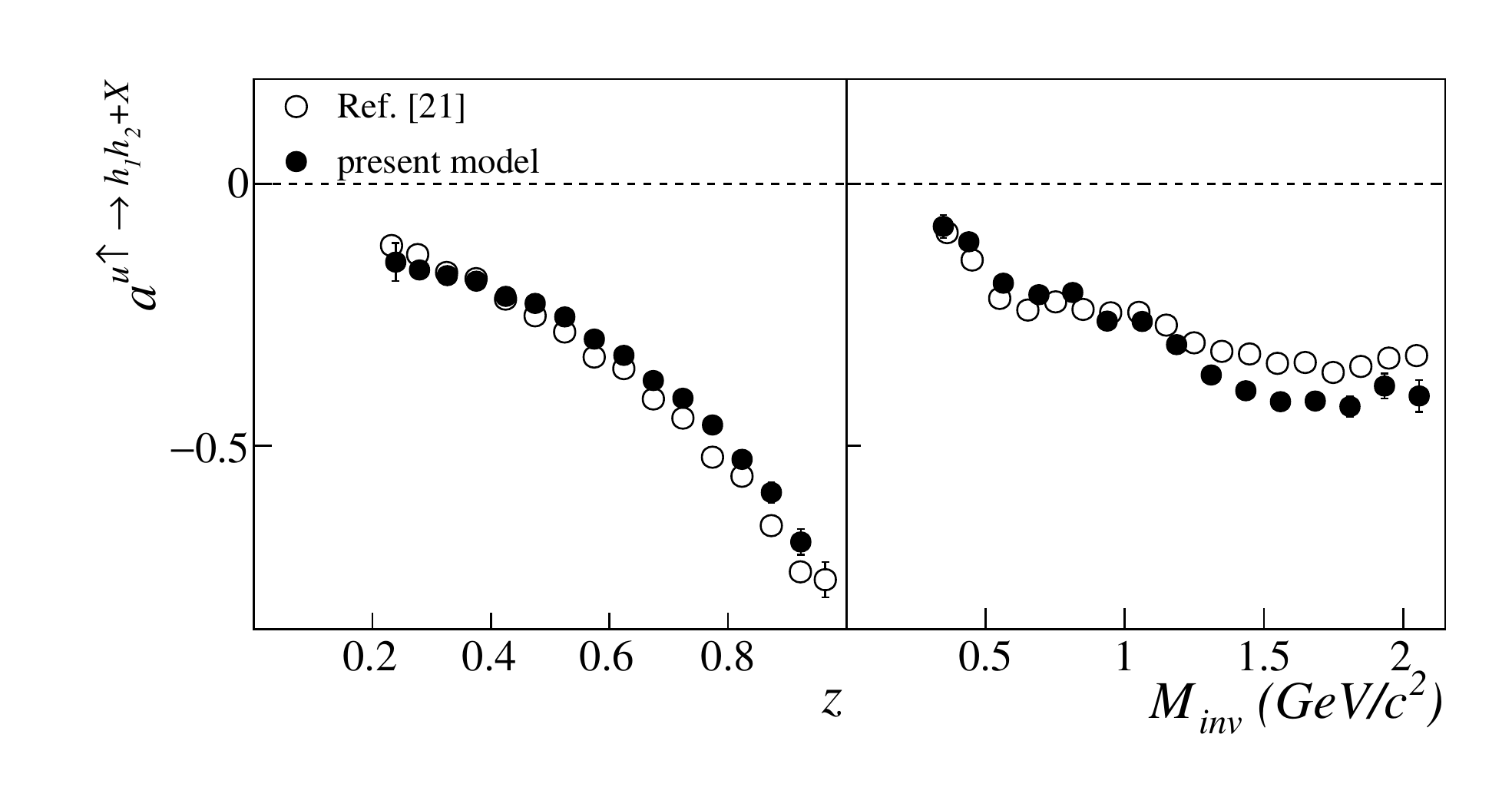}
  \caption{Comparison between the di-hadron transverse spin asymmetry as function of $z=z_{h_1}+z_{h_2}$ (left panel) and of $M_{inv}$ (right panel), as obtained for unidentified hadrons with the present model (full points) and with the model of Ref. \cite{kerbizi-2018} (open points).}\label{fig:di-hadron_asymmetry}
\end{figure}

Figure \ref{fig:di-hadron_asymmetry} compares the di-hadron $h^+h^-$ analysing power as function of $z$ (left plot) and $M_{inv}$ (right plot) as obtained with the present model (full points) and with the model of Ref. \cite{kerbizi-2018} (open points). The cuts $z_{h_i}>0.1$, $R\T>0.07\,GeV$ and $|\textbf{p}_i|>3\,GeV$ (i=1,2) have been applied.
The overall trends are the same in both models and only some slight differences can be seen. In particular as function of the invariant mass the present model saturates to somewhat larger values of the analysing power at large $M_{inv}$. All in all, the main features of the results obtained from the two implementations are the same.

\section{Positivity bounds}\label{sec:positivity}
The present simplified model allows for explicit calculations of the spin transfer coefficients between the quark $q$ and $q'$ and the positivity bounds can be checked easily.

In general a fully polarized splitting function can be defined assuming the polarization of the quark $q'$ to be analyzed by an ideally efficient polarimeter which selects only the polarization vector $\check{\textbf{S}}_{q'}$ (encoded in the matrix $\check{\rho}(q')$). Then Eq. (\ref{eq:splitting-function}) is generalized to
\begin{eqnarray}\label{eq:splitting-doubly-polarized}
\nonumber  F_{q',h,q}= \rm{tr}\left[ T_{q',h,q}\, \rho(q)\, T^{\dagger}_{q',h,q} \check{\rho}(q')\right].\\
\end{eqnarray}
Here the vector $\check{\textbf{S}}_{q'}$ is imposed. At variance with the vector $\textbf{S}_{q'}$ in Eqs. (\ref{eq:S_q'T}-\ref{eq:S_q'L}), it does depend either on $\textbf{S}_{q}$ or on the involved momenta. 
When the quark spin states are projected on the axes $M=\hat{\textbf{k}}'\T$, $N=\hat{\textbf{z}}\times \hat{\textbf{k}}'\T$ and $L=\hat{\textbf{z}}$, the fully polarized splitting function of Eq. (\ref{eq:splitting-doubly-polarized}) can be written as
\begin{eqnarray}\label{eq:F_doubly_polarized}
 \nonumber && F_{q',h,q}(Z,\pt,\check{\textbf{S}}_{q'};\kt,\textbf{S}_q)= \frac{|C_{q',h,q}|^2}{\sum_H |C_{q',H,q}|^2} \\
\nonumber &&\times \left(\frac{1-Z}{\varepsilon_h^2}\right)^a \frac{\exp{(-\bl \varepsilon_h^2/Z)}}{N_a(\varepsilon_h^2)}\\
  &&\times \frac{|\mu|^2+\kptkpt}{\langle |\mu|^2+\kptkpt\rangle_{\T}} f_{\T}^2(\kptkpt) \times \frac{1}{2}\,C(\textbf{S}_q,\check{\textbf{S}}_{q'}).
  \end{eqnarray}
The function $C(\textbf{S}_q,\check{\textbf{S}}_{q'})$ is decomposed as 
 \begin{eqnarray}\label{eq:C(Sq,Sq')}
\nonumber C(\textbf{S}_q,\check{\textbf{S}}_{q'})&=& 1+C_{N0}S_{qN}+C_{0N}\check{S}_{q'N}\\
\nonumber  &&+C_{NN}S_{qN}\check{S}_{q'N}+C_{MM}S_{qM}\check{S}_{q'M}\\
 \nonumber  &&+C_{ML}S_{qM}\check{S}_{q'L}+C_{LM}S_{qL}\check{S}_{q'M}\\
  &&+C_{LL}S_{qL}\check{S}_{q'L},
\end{eqnarray}
with $|C_{ij}|<1$, where $i,j$ take the values $M,N,L$ or $0$ in the unpolarized case.
Only the coefficients appearing in Eq. (\ref{eq:C(Sq,Sq')}) are allowed by parity conservation and are given by
\begin{eqnarray}\label{eq:Cij}
C_{N0}&=&-\frac{2\IM{\mu}\,\rm{k'_T}}{|\mu|^2+\kptkpt}=-C_{0N}\\
C_{NN}&=&-1\\
C_{MM}&=&\frac{-|\mu|^2+\kptkpt}{|\mu|^2+\kptkpt}=-C_{LL}\\
C_{ML}&=&-\frac{2\RE{\mu}\,\rm{k'_T}}{|\mu|^2+\kptkpt}=C_{LM}.
\end{eqnarray}
These coefficients describe the dynamics of the transfer of polarization from $q$ to $q'$ in the elementary splitting and are connected to the polarization vector of $q'$, given in Eqs. (\ref{eq:S_q'T}-\ref{eq:S_q'L}), through the relation
\begin{equation}
    \textbf{S}_{q'}=\frac{\nabla_{\check{\textbf{S}}_{q'}} C(\textbf{S}_q,\check{\textbf{S}}_{q'})} {C(\textbf{S}_q,\textbf{0})}.
\end{equation}
In addition they must obey the positivity conditions \cite{X.A_et_al_spin_observables}
\begin{eqnarray}\label{eq:positivity_conditions}
\nonumber (1\pm C_{NN})^2\geq && (C_{0N}\pm C_{N0})^2 + (C_{LL}\pm C_{MM})^2 \\
&&+ (C_{LM}\mp C_{ML})^2.
\end{eqnarray}
In the present model they saturate these inequalities as expected for a quantum mechanical model of the fragmentation process formulated at the amplitude level. The saturation comes from the fact that the spin-0 mesons do not carry spin information. This ensures also that the present model can be safely implemented in Monte Carlo simulations. The same is true also for the model of Ref. \cite{kerbizi-2018}, the correlation coefficients of which have more complicated expressions due to the non-vanishing $\hat{u}_{1q}(\ktkt)$ function.

\section*{Conclusions}
We have presented a new version of the string+${}^3P_0$ model of Ref. \cite{kerbizi-2018}. It is the same model but with a different choice for the input function $\check{g}$, it has the same parameters and it gives nearly the same results. The present choice of $\check{g}$ is favoured because it allows to take more simply into account the exact Left-Right symmetry and simplifies analytical calculations, as well as the extension of the model itself. The model as presented here is also more suitable to be interfaced with external event generators and in particular with PYTHIA \cite{kerbizi-lonnblad} and will also be extended to include the production of vector mesons.

\section*{Acknowledgement}
We thank Prof. Franco Bradamante for the useful discussions and comments.
\clearpage
%

\end{document}